\documentclass[aps,pra,twocolumn,groupedaddress,floatfix,showpacs]{revtex4-2}
\usepackage{amssymb,amsmath,amsthm}
\usepackage{mathtools}
\usepackage{graphicx}
\usepackage[english]{babel}
\usepackage{float}
\usepackage[svgnames]{xcolor}
\usepackage{blindtext}
\usepackage{bm}
\usepackage{comment}
\usepackage{etoolbox}
\usepackage{tikz,pgf}
\usepackage{soul}

\usepackage[symbol]{footmisc}

\usetikzlibrary{backgrounds,decorations.pathreplacing,arrows,decorations.pathmorphing,decorations.shapes,shapes.geometric,plotmarks}

\DeclareMathOperator{\arcsinh}{arcsinh}

 \newcommand{\ket}[1]{\left|#1\right\rangle}
 \newcommand{\bra}[1]{\left\langle#1\right|}

\begin{document}

\title{Transport of non-classical light mediated by topological domain walls\\ in a SSH photonic lattice}

 \author{Gabriel O'Ryan $^{1,2}$}
 \author{Joaquín Medina Dueñas$^{3,4}$}
 \author{Diego Guzmán-Silva$^{1,2}$}
 \author{Luis E. F. Foa Torres$^{1}$}
 \author{Carla Hermann-Avigliano$^{1,2}$} 
\thanks{Corresponding author: carla.hermann@uchile.cl}
\affiliation{$^1$Departamento de F\'{\i}sica,  Facultad de Ciencias F\'isicas y Matem\'aticas, Universidad de Chile, Santiago, Chile}
\affiliation{$^2$Millennium Institute for Research in Optics (MIRO), Chile}

\affiliation{$^3$ ICN2 - Institut Català de Nanociència i Nanotecnologia, Campus UAB, 08193 Bellaterra, Barcelona, Spain.}
\affiliation{$^4$ Department of Phyics, Universitat Aut{\'o}noma de Barcelona (UAB),Campus UAB, Bellaterra, 08193 Barcelona, Spain.}
\pacs{}

\begin{abstract}
Advancements in photonics technologies have significantly enhanced their capability to facilitate experiments involving quantum light, even at room temperature. Nevertheless, fully integrating photonic chips that include quantum light sources, effective manipulation and transport of light minimizing losses, and appropriate detection systems remains an ongoing challenge. Topological photonic systems have emerged as promising platforms to protect quantum light properties during propagation, beyond merely preserving light intensity. In this work, we delve into the dynamics of non-classical light traversing a Su-Schrieffer-Heeger photonic lattice with topological domain walls. Our focus centers on how topology influences the quantum properties of light as it moves across the array. By precisely adjusting the spacing between waveguides, we achieve dynamic repositioning and interaction of domain walls, facilitating effective beam-splitting operations. Our findings demonstrate high-fidelity transport of non-classical light across the lattice, replicating known results that are now safeguarded by the topology of the system. This protection is especially beneficial for quantum communication protocols with continuous variable states. Our study enhances the understanding of light dynamics in topological photonic systems and paves the way for high-fidelity, topology-protected quantum communication. 
\end{abstract}

\maketitle

\section{Introduction}

Photonics plays a fundamental role in a wide array of technological applications, including long-distance telecommunications \cite{addanki_review_2018} and precision laser-based sensors \cite{shahbaz_concise_2023}. It serves as the backbone for systems that rely on light as the medium for information transmission.
Recent years have witnessed significant advancements in the field of quantum photonics \cite{moody_2022_2022}. These systems harness photons' capacity to retain quantum properties, even at room temperature \cite{flamini_photonic_2018}, resulting in minimal noise levels and limited susceptibility to environmental interference. These advantages hold great promise for applications in communication, computation, and simulation, particularly 
using quantum light \cite{aspuru-guzik_photonic_2012,obrien_optical_2007}.
Nevertheless, challenges like error accumulation, decoherence, and the lack of precise control of quantum states present substantial hurdles, particularly when striving for secure, stable, and long-distance quantum computing.

Waveguides arrays emerge as a dependable, high-speed, and cost-effective platform to transport radiation \cite{davis_writing_1996,nasu_low-loss_2005}. Light is confined to travel within individual waveguides, with a probability of transitioning to their neighbor guides depending on different parameters such as geometry and index of refraction. The challenges encompass dispersion and losses during propagation and inefficient light coupling to the waveguides. These are highlighted when the objective is to preserve quantum properties of light \cite{slussarenko_photonic_2019}. 

Topological systems may be considered as a natural scenario to propagate quantum light more robustly, since it offers robustness in comparison to topologically trivial arrays \cite{hafezi_imaging_2013}.
 Topology in a photonic array influences the quantum characteristics of light, extending beyond the qualities of localization and robustness often associated with classical light (intensity). Quadrature protection of squeezed light and enhanced teleportation protocols have been proposed \cite{duenas_quadrature_2021}, which is remarkably important when performing quantum information protocols with continuous variables \cite{braunstein_quantum_2005}. 
%%%%%%%%
\\ \indent In this work, we study a Su-Schrieffer-Heeger (SSH) photonic lattice with domain walls (DW). This array has been deeply studied due to its simplicity yet clear topological nature, making it a suitable candidate to study topological effects on new platforms. The concept of a domain wall is widely employed to denote the boundary between two different behaviors \cite{kumar_domain_2022}. In our context, when two systems in a different topological phase are connected, a domain wall is created on the junction and the number of states on it is determined by the difference between topological invariants \cite{asboth_short_2016}. 
We re-positioned the DWs across the lattice through a local modulation of the coupling between waveguides. Light injected into the DW follows its movement. Moreover, we can approach the DWs as much as needed, until 
they start to interact, forming effective couplers such as a dimer.
The system is fine-tuned to minimize dispersion during modulation, which prevents the loss of the initial state. Our findings show how topological properties bestow robustness to quantum light while it propagates between DWs.
%%%%%%%%%%%%%%%%%%%%%%%%%%%%%%%%%
%%%%%%%%%%%%%%%%%%%%%%%%%%%%%%%%%
\section{ SSH lattice with domain walls}
\begin{figure}
    \includegraphics[width=8.6cm,height=12.5cm]{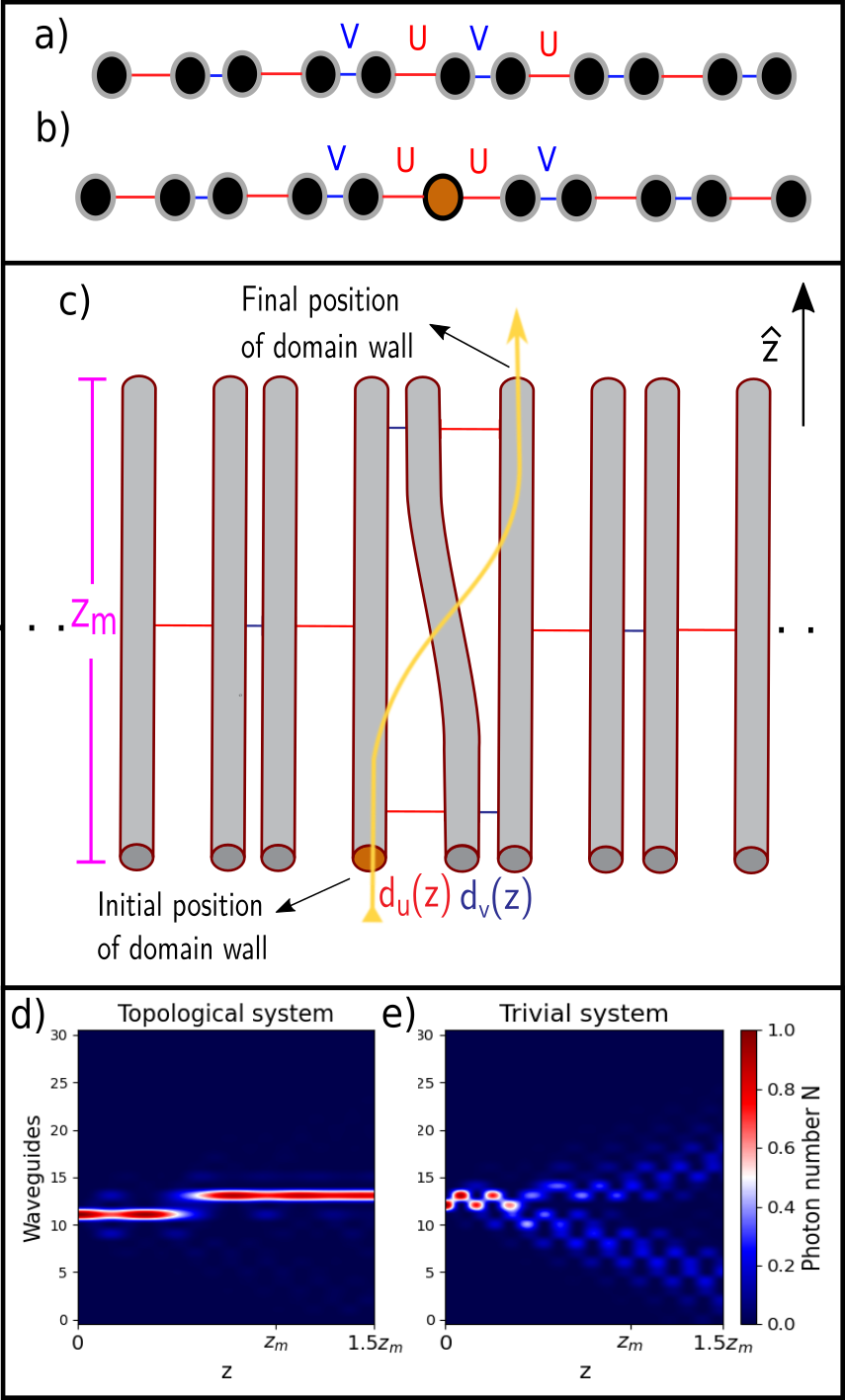}
    \caption{\textbf{a)} SSH array with intra-coupling $u$ and inter-coupling $v$. \textbf{b)} SSH array with a topological domain wall. \textbf{c)} Example of how the bending is done. The distance between waveguides depends on the propagation and the whole process has a distance of $Z_{m}$ called the modulation length. This is the principal mechanism to transport light across our lattice. \textbf{d)} Shows the photon number evolution through the bending in a topological domain wall. \textbf{e)} Shows the photon number evolution through the bending in a trivial domain wall. }
    \label{Resumen_setup}
\end{figure}

The SSH model \cite{asboth_short_2016,su_solitons_1979} consists of a one-dimensional dimerized chain, as depicted in Figure \ref{Resumen_setup}a). The unit cell possesses two sites with intra-coupling $u$ and inter-coupling $v$. Defining $\delta=|v/u|$, if $\delta>1$, the system is in a topological phase, and two zero-energy states appear within the energy gap. These states are exponentially localized on the edges, possess sub-lattice symmetry, and are robust to coupling disorder \cite{asboth_short_2016}.
 By either repeating coupling $u$ or $v$ inside the array, a domain wall is added - which is equivalent to joining two topological systems with different edges - as depicted in Fig. \ref{Resumen_setup} b). The Hamiltonian of the system is
%%%%%%%%%%%%%%%%%%%%%%%%%%%%
\begin{equation}
\hat{\mathcal{H}}  =\sum_{n\neq m} (u\hat{a}^\dagger_n\hat{b}_n+v\hat{b}^\dagger_n \hat{a}_{n+1}+\delta_{n,m}u\hat{a}^\dagger_n\hat{a}_{n+1})+ h.c.\;,
\end{equation}
%%%%%%%%%%%%%%%%%%%%%%%%%%
where $a^\dagger_n(a_n)$ and $b^\dagger_n(b_n)$ are the creation (annihilation) operator at site $n$ of the corresponding unit cell and it is assumed there is only one domain wall in site $m$ with coupling $u$. Although either coupling $u$ or coupling $v$ can be used in the interface to create a domain wall, we will study the one with coupling $u$, because it allows having an isolated state on the domain wall. 
The DW state shares the topological properties of the edge state and it hybridizes with the nearest edge if it is in a topological phase. This provokes different dynamics compared to a domain wall in a trivial phase as seen in Fig. \ref{Resumen_setup} d,e). The propagation of light inside both types of domain wall was studied in \cite{blanco-redondo_topological_2016}.
%%%%%%%%%%%%%%%%%%%%%%%%%%%%%%%%%%%%%%%%%%%%%%%%%%%%%
%%%%%%%%%%%%%%%%%%%%%%%%%%%%%%%%%%%%%%%%%%%%%%%%%%%%%
%%%%%%%%%%%%%%%%%%%%%%%%%%%%%%%%%%%%%
\subsection{Guiding non-classical light through topological domain walls}
%%%%%%%%%%%%%%%%%%%%%%%%
Given the Hamiltonian in Equation (1), it is possible to find the infinitesimal unitary operator $\hat{U}(z+dz)$ which can be used to obtain the entire evolution, even for time-dependent Hamiltonians (see Appendix \ref{Numerical_method}). The operator $\hat{U}$ depends on the propagation distance $z$ (analog to time in condensed matter systems). Here we explore the properties of the array when propagating quantum states of light. We consider four initial states $\ket{\psi_{0}} $ in the DW: single photon, coherent, vacuum single-mode squeezing, and vacuum two-mode squeezing states. As pure states, they can be described as operators acting on the vacuum \cite{scully_quantum_1997}
\begin{align} \label{initial_states}
& \ket{\psi_{0}} = \hat{a}^{\dagger}_{DW} \ket{0} = \ket{1} \\
& \ket{\psi_{0}} = \hat{D}_{DW}(\alpha)\ket{0} = \ket{\alpha} \\
& \ket{\psi_{0}} = \hat{S}_{DW}(\xi) \ket{0} = \ket{\xi} \\
& \ket{\psi_{0}} = \hat{S}_{0,DW}(\xi) \ket{0} = \ket{\xi_{0,DW}}\;,
\end{align}
where $\hat{a}^{\dagger}_{DW}(\hat{a}_{DW}) $ is the creation (annihilation) operator on the position of the DW, $\hat{D}_{DW}(\alpha) = \exp{(\alpha \hat{a}_{DW}^{\dagger} - \alpha^{*}\hat{a}_{DW})}$ the displacement operator on the DW  , $\hat{S}_{DW}(\xi) = \frac{1}{2}[\xi^{*} \hat{a}_{DW}^{2} - \xi (\hat{a}_{DW}^{\dagger})^{2}]$ the single-mode squeezed operator on the DW and $ \hat{S}_{0,DW}(\xi) = \frac{1}{2}(\xi^{*} \hat{a}_{DW}\hat{a}_{0} - \xi \hat{a}_{DW}^{\dagger}\hat{a}_{0}^{\dagger}) $ the two-mode squeezed operator between the edge and the DW. We choose the edge as the other half because the edge state has the same energy as the DW state. The amplitude of the states is computed by taking the mean value of the photon number operator
%%%%%%%%%%%%%%%
\begin{equation}
    \langle\hat{N}_{i,j}(z)\rangle = \bra{\psi(z)}\hat{a}_{i}^{\dagger}\hat{a}_{j}\ket{\psi(z)}\;.
\end{equation}
%%%%%%%%%%%%%%%%%%%
To study squeezed light, we use the usual decibel function that compares the state's uncertainty with the one of a coherent state,
\begin{equation} \label{Squeezing_formula}  
    S_{i}^{(1,2)}(z,\phi)[\text{dB}] = 10 \log_{10} \Big(  \frac{\bra{\psi(z)}\Delta \hat{X}^{(1,2)}_{i}(\phi)\ket{\psi(z)}^{2}}{\langle \Delta \hat{X}^{(1,2)}_{i} \rangle^{2}_{coh}}    \Big)\;,
\end{equation}
with $\langle\Delta \hat{X}\rangle^{2} = \langle \hat{X}^{2}\rangle - \langle \hat{X} \rangle^{2}$ the standard deviation. For this we consider the generalized quadrature as
\begin{equation}
    \hat{X}_{i}(\phi) = \frac{1}{2}(\hat{a}_{i}e^{i\phi} + \hat{a}^{\dagger}_{i} e^{-i\phi} )\;.
\end{equation}
The two orthogonal quadratures are then $\hat{X}^{(1)}_{i}(\phi) = \hat{X}_{i}(\phi)$ and $\hat{X}^{(2)}_{i}(\phi) =\hat{X}_{i}(\phi + \frac{\pi}{2})$ and the two-mode quadrature is $\hat{X}^{(1,2)}_{ij} = \frac{1}{\sqrt{2}}\big( X^{(1,2)}_{i}(\phi) + X^{(1,2)}_{j}(\phi)\big)$, where $\phi$ is the phase of the quadrature.  We can study the dynamics of the squeezing by computing the uncertainty of both orthogonal quadratures.
 For simulations and comparisons between the states, we set the mean photon number equal to 1, meaning that for the coherent state $\alpha =1$ and for the squeezed state $\xi =\arcsinh (1)$ (7.65 [dB] in squeezing magnitude). We observe that the light propagates in the domain wall only when the array is in the topological phase (see Figure \ref{Resumen_setup}d)). If we make a DW using the coupling $v$ and then switch the couplings so that $\delta < 1$, we get a DW between coupling $u$ but in a trivial phase. In this trivial case,  the initial input just spreads along the waveguides if we try to transfer it as shown in Figure \ref{Resumen_setup} e). We propose a method to move domain walls across the lattice by slowly shifting two adjacent waveguides. In photonic lattices, this can be achieved by bending the waveguide, as shown in Figure \ref{Resumen_setup}c). This moves the domain wall to the next adjacent waveguide, either to the left or right, depending on the direction of curvature. The objective is to guide the light and transport it to its new position with minimum dispersion, which depends on how we perform the bending (how fast, how sharp).
To present our system as an experimentally feasible one, we assume realistic values and limits. The coupling $C$ is related to the distance $D$ between waveguides as $C = c_{2} \exp(- c_{1}D)$, with the constants $c_{1,2}$ depending on the parameter of fabrication \cite{szameit_control_2007}. The main parameters to consider are the coupling ratios $\delta$, the modulation length $Z_{m}$, and the slope $s$ of the bending. Logistic function or s-shape functions are usually picked for activation or adiabatic processes due to their expected suppression of leakage \cite{knapp_nature_2016,wiebe_improved_2012}. 
Among this family, we pick the function
%%%%%%%%%%%%%%%
\begin{equation}
    f(z) =A- \frac{B \exp(-Z_{m}/z)}{  s \exp[-1/(1-z/Z_{m})] + \exp(-Z_{m}/z)}\;.
\end{equation}
%%%%%%%%%%%%%%%%
A slight variation is used in \cite{boross_poor_2019} resulting in low leakage. The constants $A$ and $B$ are obtained by imposing the initial coupling difference while $s$ controls the slope. The $s$ parameter is optimized to give the greatest transmission given the coupling difference and modulation length. It also serves a direct variable to change the radius of curvature.
%%%%%%%
%%%%%%%%%%%
While many parameters fit the boundary conditions, not all of them are realistic. Previous works show that bending waveguides introduce radiation losses proportional to the radius of curvature, meaning that a sharper bend generates more losses \cite{eaton_telecom-band_2006,calmano_curved_2013}. This imposes a limitation if the curved waveguide is guiding light. Remarkably, our model does not have this limitation, because curved waveguides only serve as a bridge between straight waveguides, allowing to spatially move each DW.
For the following sections, we consider a lattice of $32$ waveguides and parameters (see Figure \ref{Resumen_setup}) $d_{u} = 22~\mu$m, $d_{v} = 10 ~\mu$m , $Z_{m} = 5.5$~cm and $ s = 1.5$. The couplings used are $ u = 0.69~\text{cm}^{-1} $ , $ v = 3.22~\text{cm}^{-1} $ and $ \delta = 4.62 $. More details about this lattice can be found in Appendix \ref{system_properties}.
%%%%
\begin{figure}
    \centering
    \includegraphics[width=8.6cm,height=11cm]
    {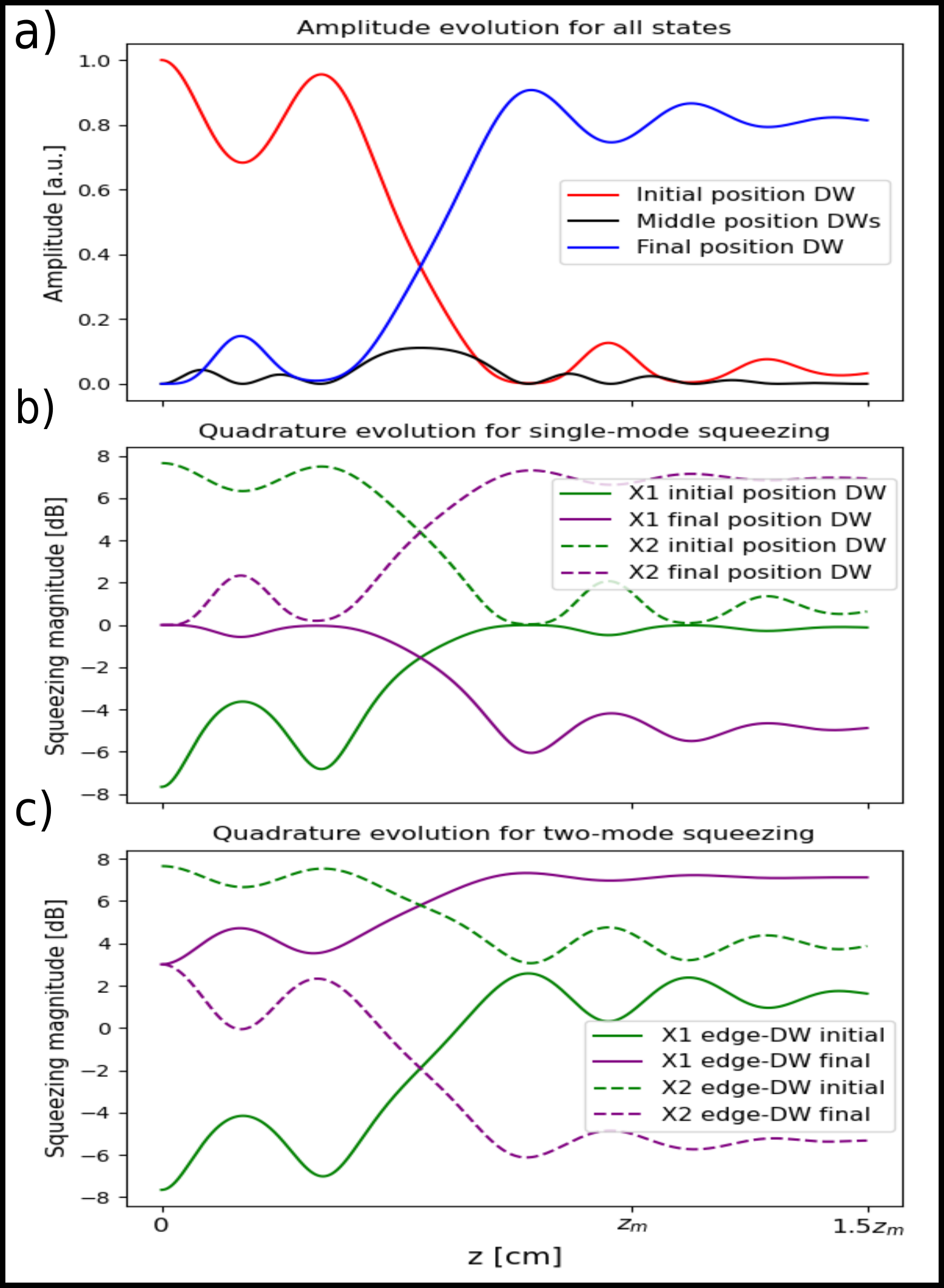}
    \caption{\textbf{a)} Shows the photon number evolution for the four initial states (equations 2 to 5) across the waveguides, from the initial to the final position of the DW as depicted in Fig. \ref{Resumen_setup}c). \textbf{b)} Shows the single-mode quadrature magnitude evolution of the single-mode squeezed state when moving the DW. \textbf{c)} Shows the two-mode squeezed quadrature magnitude evolution between the edge and the moving DW.}
    \label{single_move_evolutuion}
\end{figure}

%%%%%%%%%%%%%%%%%%%%%%%%%%%%%%%%%%%%%%%%%%
\begin{figure*}[t!]
    \centering
    \includegraphics[width=\textwidth]{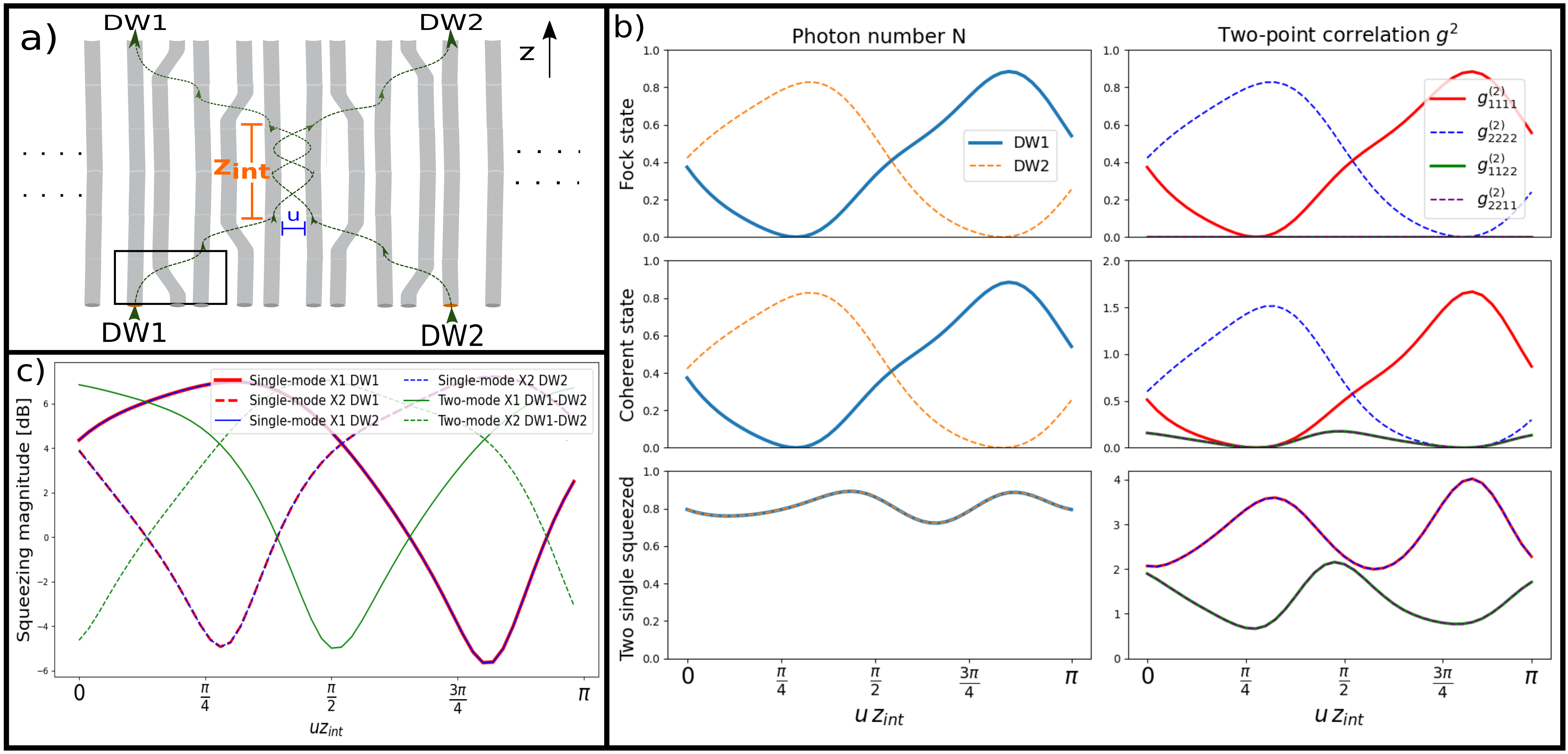}
    \caption{\textbf{a)} Example of the effective beam-splitter. The black box encapsulates the size of one movement. Light is injected on DW1 and DW2, interferes at the middle for a length $z_{int}$ with coupling $u$, and then returns to the same waveguide that was injected on. \textbf{b)} Photon number $N$ (left column) and two-point correlation function $g^{2}$ (right column) versus normalized distance $uz_{int}$ for three different input states: Single photon Fock state in DW1, coherent state of amplitude $\alpha=1$ in DW1 and two single-mode squeezed states of amplitude $\xi=\arcsinh (1)$ in both DWs. \textbf{c)} Shows the evolution of single and two-mode quadratures within and between DWs, respectively.}
    \label{n2_g2_figure}
\end{figure*}
%%%%%%%%%%%%%%%%%%%%%%%%%%%%%%%%%%%%%%%%%%
The photon number evolution for all initial states is shown in Fig. \ref{single_move_evolutuion}a). All states show the same $N(z)$ evolution, which is expected in waveguides systems because $N$ evolves in the same way as intensity distribution of classical light \cite{rai_transport_2008,bromberg_quantum_2009} and we have treated it as a closed system. For all the states we considered, $\sim80\% $ of the initial photon number N could be transported. The 20\% remaining is scattered during dynamics due to dispersion to other parts of the system.
If the bending is done optimally, the light disperses to the same sublattice as the DW and nothing enters the bent waveguide. After the transport, 
the light remains localized on the new position. Current studies of transport within waveguide lattices have shown that the initially squeezed quadrature rotates \cite{swain_non_2021} proportional to the transported distance, not including disorder effects. In contrast, in our system, the phase of the initially single-mode squeezed quadrature is maintained during the dynamics as shown in Fig. \ref{single_move_evolutuion}b). The two-mode squeezing presents a similar protection, but in this case, the squeezed quadrature rotates by $\pi/2$ and the correlation with the edge is transported to the new DW. This is due to a phase difference generated by moving the DW but not the edge state. Moving two DWs with a two-mode squeezing between them does not generate this rotation.   
Then, with this procedure, we can transport non-classical light across two waveguides with transmission higher than 80\%. This can be repeated to move the DW to a new position of the same sub-lattice. Just like the amplitude, the initially squeezed quadrature remains localized and locked after the transport ends.

%%%%%%%%%%%%%%%%%%%%%%%%%%%%%%%%%%%%%%%%%%%%%%%%%%%%%%%
\section{Topological beam splitter}
%%%%%%%%%%%%%%%%%%%%%%%%%%%%%%%%%%%%%%%%%%%%%%%%%%%%%%%

One interesting point to observe in topological arrays is the interaction between two (or more) topological states since this interaction can perform a primary element of any photonic circuit as it is a beam splitter operation \cite{tambasco_quantum_2018}. In our scheme, it is possible to obtain an effective beam splitter with two DW (or more for a multiport splitter), as depicted in Figure \ref{n2_g2_figure} a). Light is injected into the moving DWs splitting the signal. After a fixed length, both domain walls are separated one more time into their original waveguides (not mandatory, other waveguides can be used), and the output is then studied. To demonstrate the splitting and transport properties we propagate a coherent state in one DW and vacum in the other $\ket{\alpha}\otimes \ket{0}$, a Fock state $\ket{1}\otimes \ket{0}$, and two single mode squeezed states $\ket{\xi}\otimes \ket{\xi}$ as different input conditions. We analyze the output for different lengths of interaction $z_{int} $, taking in consideration the photon number $N$ and the correlation function $g^{(2)}$, both depending on the normalized interaction length
\begin{align}
& \hat{N}_{i,j}(u z_{int}) = \langle\hat{N}_{i,j}\rangle = \langle\hat{a}_{i}^{\dagger}\hat{a}_{j}\rangle(u z_{int}) \label{photon_number} \\
& g^{(2)}_{i,j,k,l}(u z_{int}) = \langle\hat{N}_{i,j}\hat{N}_{k,l}\rangle(u z_{int}) = \langle\hat{a}_{i}^{\dagger}\hat{a}_{j}\hat{a}_{k}^{\dagger}\hat{a}_{l}\rangle(u z_{int}) \label{g2_function}\;.
\end{align}
%%%%%%%%%%%%%%%%
We observed that the output display in Figure \ref{n2_g2_figure} b), is analogous to the propagation inside two waveguides (dimer) for the three cases studied here \cite{rodriguez-lara_propagation_2014,rojas-rojas_manipulation_2019}.
%%%%%%%%%%%%%%%%%%%%%%%%%%%%%%%%%%%
\begin{figure*}
    \centering
    \includegraphics[width=\textwidth]{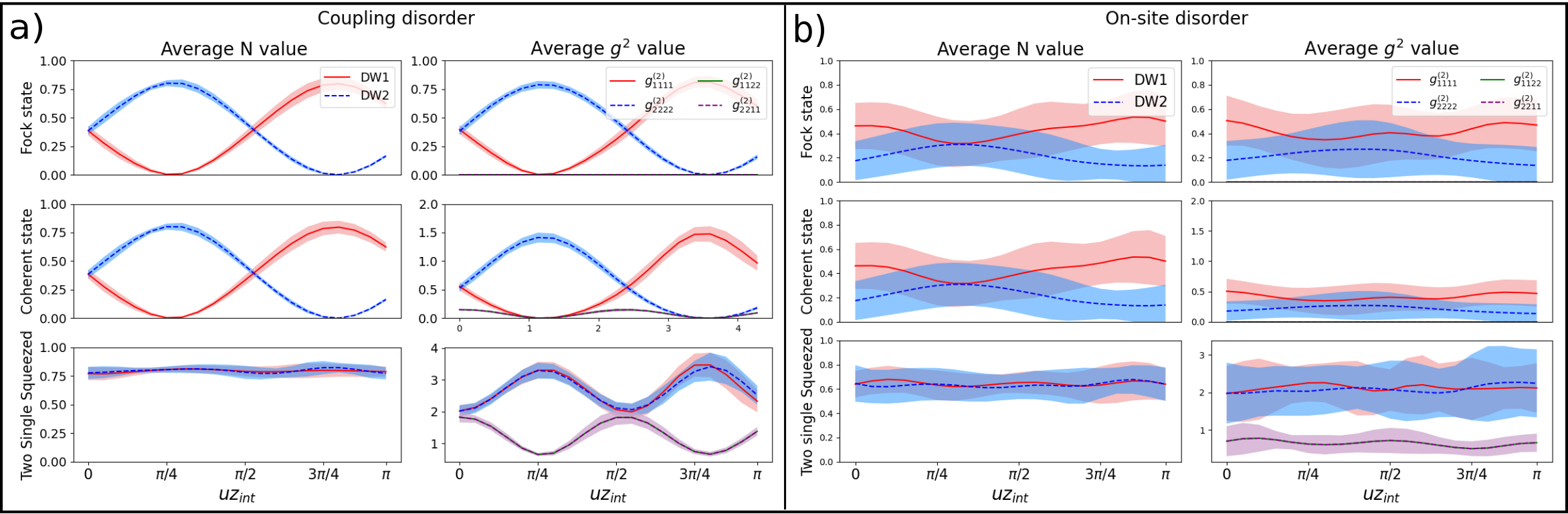}
    \caption{ \textbf{a)} Show the average  (continuous and dashed lines) and standard deviation (shaded area) for 20 repetitions of the output photon number and two-point correlation function on each DW output as depicted in Fig. \ref{n2_g2_figure}b) when it is subject to a uniform coupling disorder $\Delta=\pm1.3$ cm$^{-1}$.
    \textbf{b)} Same as in part a) but for the case of on-site disorder.}
    \label{desorden_n_g2}
\end{figure*}
%%%%%%%%%%%%%%%%%%%%%%%%%%%%%%%%%%%%%%%%%%%%%%%%%%
The single photon state oscillates between DWs for both $N$ and $g^{2}$ functions, not generating a correlation between waveguides.  The coherent state with $\alpha=1$ behaves similarly but has a higher value in the correlation function (diagonal) and presents low off-diagonal correlation values between the DWs at $u z_{int} = (0,\frac{ \pi}{2})$. This is because for the coherent state $\ket{\alpha} = e^{-|\alpha|^{2}/2} \sum_{0}^{\infty} \alpha^{j}(j!)^{-1/2}\ket{j} $, the initial vacuum component does not transfer between waveguides. Hence, for a coherent state with $\alpha \leq |1|$, the vacuum amplitude is comparable to the other amplitudes, thus the appearance of off-diagonal correlations \cite{rodriguez-lara_propagation_2014}. This correlation becomes negligible as the photon number increases. The squeezed states present a constant photon number due to being injected on both DWs. The decrease in the diagonal correlation while increasing the off-diagonal, shows the formation of two-mode squeezing around $u z_{int} = (0,\frac{\pi}{2})$ as shown in Fig. \ref{n2_g2_figure}c). The states go back to being single mode squeezed light at around $u z_{int} = (\frac{\pi}{4}, \frac{3\pi}{4})$. This behavior oscillates as a function of $uz_{int}$, just as how a dimer should behave with such a non-classical light \cite{rojas-rojas_manipulation_2019}.

By using DWs we can transport non-classical light across the system and by the interaction of DWs, light can be split in different waveguides. The transmission stays above 80\% which is a bit larger than one DW movement transmission. This is because in the case of only one DW, part of the light is dispersed into the array (there is no perfect transmission). In contrast, having two DWs allows the leakage of one DW to be ``caught" by the other and vice versa. We also emphasize that the energy band never closes and the system is always in the same topological phase along the whole dynamics. Besides from photon number, the topology protects the initially squeezed quadrature, and transporting the state does not rotate it. This prevention of rotation is crucial as both single-mode states entering the effective dimer must have the same phase to maximize the generation of two-mode squeezed light. Rotation of the quadrature and formation of two-mode squeezing starts just when the effective dimer forms. 

%%%%%%%%%%%%%%%%%%%%%%%%%%%%%%%%%%%%%%

\subsection{Robustness against disorder}
%%%%%%%%%%%%%%%%%%%%%%%%%%%%%%%%%%%%%%

We test our system against two kinds of disorders: the off-diagonal disorder, which affects the coupling between waveguides, and the diagonal disorder, which mainly depends on the size and change of refractive index of the waveguide (reflected in the on-site energy parameter). The first one is known to preserve the chiral symmetry, while the second does not. We add a uniform disorder $\Delta$ for both cases, with a value inside the interval of $[-1.3,1.3]$ cm$^{-1}$. We observe that the system is robust against coupling disorder maintaining the output for both N and $g^{(2)}$, as shown in figure \ref{desorden_n_g2} a). The squeezing state is particularly interesting, as discussed before, due to the generation of maximal two-mode squeezing (i.e. without single mode contributions) inside a dimer. Under disorder, a trivial array is likely to fail as the phase varies randomly. Nevertheless, in our system due to the topological protection the state generates almost no rotation of the quadrature preventing decoherence in the squeezed phase and allowing the formation of two-mode squeezed light.

For the case of on-site disorder, the results show a great dispersion (shaded area in Fig. 4) for both $N$ and $g^{2}$. This contrasts with the case of off-diagonal disorder, implying a break of chiral symmetry.
The $g^2$  correlation function for the squeezed states behaves noticeably worse than the Fock and coherent state even though the decrease in photon number is similar. This is because on-site disorder, apart from destroying the state, rotates the quadrature randomly before arriving at the effective dimer, causing both single-mode states to have a decoherence in the squeezed phase and unable to generate maximal two-mode states.\\

\section{Conclusion}
We presented a photonic waveguide model featuring topological domain walls capable of guiding light by recursively bending waveguides. Topology emerges as a crucial ingredient for this mechanism to work since a trivial domain wall cannot achieve the same behavior. The quality of the transport can be optimized with lattice parameters such as the propagation length and slope of the bending. We optimize our system, achieving above 80\% transmission using realistic experimental values. To test our system, we used Fock states, coherent states, and squeezed states, finding in the latter that topology prevents the initially squeezed quadrature from rotating during transport. The interaction of two domain walls forms an effective dimer acting as a beam-splitter to the injected light.  The phase protection during the transport allows the effective dimer to generate two-mode squeezing even in the presence of coupling disorder, contrasting, on-site disorder degrades the photon number and generates decoherence in the squeezed phase. Our protocol can be extended to multi-port splitting by adding more DWs.
We think that the ability to transport quantum light, particularly squeezed states, protecting the squeezed quadrature and not just its amplitude will be beneficial in different quantum communications protocols \cite{duenas_quadrature_2021}.

\bibliography{references.bib}

\appendix
    \section{Band structure and distribution of states} \label{system_properties}
\begin{figure}[h]
        \centering
        \includegraphics[width=7cm,height=9cm]{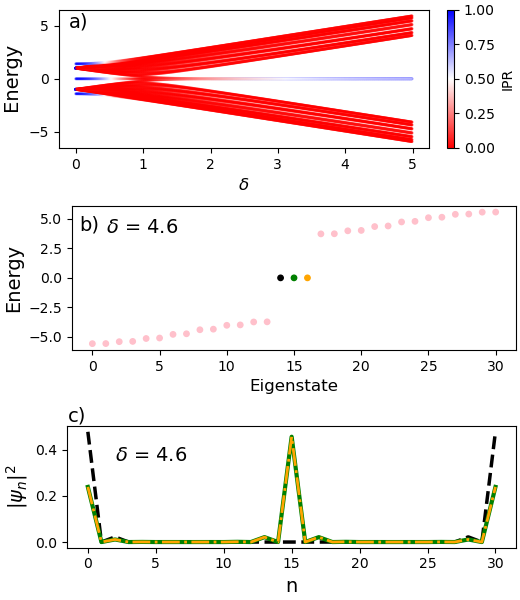}
        \caption{\textbf{a)} Energy bands as a function of the coupling relation $\delta$. The inverse participation ratio is represented with a color scale for each state. \textbf{b)} Example of the energy spectrum for $\delta = 4.6$. The topological states are highlighted. \textbf{c)} Wave-function distribution across the lattice for the highlighted zero energy states shown in b).}
        \label{DW_bands_localization}
    \end{figure}

    The energy spectrum and distribution of states for a SSH model of 31 sites hosting a domain wall at position 15 and in a topological phase is shown in Figure \ref{DW_bands_localization}. We define the inverse participation ratio as $\text{IPR} = \sum_{n}|\psi_{n}|^{4} / (\sum_{n} |\psi_{n}|^{2})^{2}$. The IPR of a localized state tends to 1 while for extended states it tends to 0.
    %%%%%%%
    For $\delta >1$ the system is in a topological phase and we only have states with low IPR inside the energy gap corresponding to the two edges and domain wall. The domain wall state decays exponential to both sides and on the same sub-lattice due to sub-lattice symmetry. The state on the domain wall hybridizes to the edges when the system is in a topological phase, this causes that part of the state to have weight on the edge, lowering the IPR.
    %%%%%%%%%%%%%%%%%%%%%%%%%%%%%%%%%%%%%%%%%%%%%%%%%%%%%%%%5
	\section{Numerical method for the evolution of quantum states of light in time-dependent linear Hamiltonian.} \label{Numerical_method}
    In order to get the expectation values for the photon number and two-point correlation function (using a time-dependent Hamiltonian), we exploit the linearity of the Hamiltonian and a time-dependent trotter algorithm \cite{poulin_quantum_2011} to compute these quantities.
    A linear Hamiltonian can be written as $\hat{\mathcal{H}} = \sum_{m,n}  H_{mn} \hat{a}^{\dagger}_{m} \hat{a}_{n}$ and, if it is independent of time, the evolution operator is $\hat{U}(t_{0},t) = \exp\{-i(t-t_{0})\hat{\mathcal{H}} \}$. In the Heisenberg picture and using the BCH formula, the ladder operators fulfill:
    %%%%%%%%%%%%%%
    \begin{equation} \label{mode_evolution}
    \hat{a}_{j}(z) = e^{ib\hat{\mathcal{H}}} \hat{a}_{j} e^{-ib\hat{\mathcal{H}}} = \sum_{n} e^{-ib H_{jn}} \hat{a}_{n} . 
    \end{equation}
    %%%%%%%%%%%%%%%%
    For time-dependent Hamiltonians we need to use the trotter approximation as:
    %%%%%%%%%%%%
    \begin{align} \label{evolution_operator_infinitesimal}
    \hat{U}(t_{0},t_{0} + \Delta t) &\approx \exp\left\{ -i \int_{t_{0}}^{t_{0}+\Delta t} \hat{\mathcal{H}}(s) ds \right\} \nonumber\\
    &= \exp\{- B(t_{0},\Delta t)_{mn} \: \hat{a}_{m}^{\dagger} \hat{a}_{n}\}\;.
    \end{align}
    %%%%%%%%%%%%%
    The evolution of the ladder operator in this case in the Heisenberg picture looks as:
    \begin{align}
        \hat{a}_{m}(t_{0} + \Delta t) &= \hat{U}^{\dagger}(t_{0},t_{0} + \Delta t) \hat{a}_{m}(t_{0}) \hat{U}(t_{0},t_{0} + \Delta t) \nonumber\\
        &= \sum_{n} U_{mn}(t_{0},\Delta t) \hat{a}_{n}(t_{0})\;,
    \end{align}
    %%%%%%
    with $U_{mn}(t_{0},\Delta t) = \exp\{-i\int_{t_{0}}^{t_{0}+\Delta t} H_{mn}(s) ds \}$, a (N,N) matrix with N the size of the lattice.
    To calculate the evolution of the expectation values for the photon number and two-point correlation function we compute:
   %%%%%%%%%%%%%%%%%%%%%%%%%%
    \begin{equation}
        \begin{split}
            N_{ij}(t_{0}+ \Delta t) &= \bra{\psi(t_{0}+\Delta t)}\hat{a}_{i}^{\dagger}\hat{a}_{j}\ket{\psi(t_{0}+\Delta t)} \\
             &= \sum_{nm} U^{*}_{im}(t_{0},\Delta t) U_{jn}(t_{0},\Delta t) N_{mn}(t_{0})\;. 
        \end{split}
    \end{equation}
%%%%%%%%%%%%%%%%%%%%%%%%%%%%%%%
    The initial condition $N(t_{0})$ is calculated analytically, then the simulation consists of repeatably multiplying the two evolution matrices with the $N_{mn}$ on each time.
    For $g_{ijkl}^{(2)}$ the recipe is the same but four contractions are needed, one for each index as:
    \begin{equation}
        g_{ijkl}^{(2)}(t_{0} + \Delta t) = \sum_{mntp} U^{*}_{im} U_{jn}  U^{*}_{kt} U_{lp} g^{(2)}_{mntp}(t_{0})\;. 
    \end{equation}
    Again, we just need to calculate the initial condition for the correlation $g^{2}$ and the evolution is computed iterating. 
    To calculate the uncertainty we use quadrature $\hat{X}_{i}^{(1)}$
    \begin{equation}
        \langle \Delta \hat{X}_{i}^{1} \rangle = \frac{1}{4}\big(\langle\hat{a}_{i}\hat{a}_{i}\rangle e^{-2i\phi} + \langle\hat{a}_{i}^{\dagger}\hat{a}_{i}\rangle +\langle\hat{a}_{i}\hat{a}_{i}^{\dagger}\rangle +\langle\hat{a}_{i}^{\dagger}\hat{a}_{i}^{\dagger}\rangle^{2i\phi}  \big)\;.
    \end{equation}
    Here we used $\langle\hat{a}_{i}\rangle_{sq} = \langle \hat{a}_{i}^{\dagger}\rangle_{sq} = 0$, because vacuum squeezed states have zero first momenta. The orthogonal quadrature is obtained analogously. Using the same method as before we can compute the evolution for all these matrices and obtain the evolution for the uncertainty, which corresponds to Eq. (\ref{Squeezing_formula}) of the main text.

\end{document}